\documentclass[12pt]{article}
\usepackage{graphicx}
\usepackage{cite}
\usepackage{amssymb,amsmath}
\begin{document}
\title{The effect of DNA conformation changes on the coupling of the macromolecule deformation components}%
\author{P.P. Kanevska, S.N. Volkov\\
Bogolyubov Institute for Theoretical Physics, NAS of
Ukraine,\\14-b Metrolohichna Str., Kiev 03143, Ukraine \\
snvolkov@bitp.kiev.ua }\maketitle
\setcounter{page}{1}%
\maketitle
\begin{abstract}
The model of the deformation of DNA macromolecule is developed
with the accounting of two types of components of deformation:
external and internal. External components describe the bend,
twist and stretch of the double helix. The internal component -
the conformational mobility inside of the double helix. In the
work the deformation of DNA macromolecule is considered taking
into account the coupling of the external component (of
deformation) with the internal component (of conformational
change). Under the task consideration the macromolecule
twist-stretch coupling and coupling between twist and internal
component are taken into account. The solution obtained in these
conditions for the deformation components allows changing the
character of respond in stretch component on unwind (overwind) in
dependence on the applied force to twist component.

The changing of the character of deformation from compression to
tension achieving of critical untwisting force (and vise versa the
changing of the character of deformation from overwind to unwind
at critical tension force)is known from the single molecular
experiments   \cite{Ref24,Ref26,Ref26a}. The nature of such
unexpected behavior of double helix have clarified in the present
work by including in consideration the internal component. The
obtained solutions and their conformity to experimental results
show the essential role of coupling between internal and external
components in the double helix conformational mechanics under
action force in pN range.
\end{abstract}
\section{Introduction}
Determination of the deformation mechanism of DNA double helix is
an important task for understanding the functioning of the
macromolecule \cite{Ref6,Ref3,Ref22}. For a theoretical
description of DNA double helix deformations, the model of the
elastic rod or worm-like chain (WLC) model are commonly used.
These models describe a macromolecule as infinitely long chain of
homogeneous monomer links and the deformation of macromolecule is
considered as the small displacement between adjacent links. The
displacements are usually considered using the independent
external components: bend, twist, stretch within the framework of
the elastic rod model \cite{Ref1,Ref25}.

However, due to the DNA helical structure the deviations from
equilibrium state usually accompany by the change in two or even
all of three components. The single molecule experiments allow to
measure twist under manipulating with stretching and vice versa
stretch during twist manipulating \cite{Ref24,Ref26,Ref26a,Ref33},
that confirms the existence of twist-stretch coupling. In the case
of stretching force lower then critical value (about $~35pN$) the
stretching induces the winding of the double helix (positive
twist), while in the case of the force higher than critical one
the stretching induces the unwinding of the double helix (negative
twist).

This unexpected behavior have been modelled with the changing of
coupling parameter under the force \cite{Ref33,Ref33a}. In the
same time such changing could be induced by coupling with
component associated with intrinsic conformational mobility of
double helix. The approach allowing extract internal component for
describing deformation of double helix during conformational
transitions of $\mathcal{B-A}$ type was developed by S.N.Volkov
\cite{Ref27,Ref28,Ref29}. This approach is appropriate for
deformations accompanied with changing in internal conformation of
double helix and could be generalized for deformations with
coupled external components such as twist, stretch and bend.

 Really, the molecular dynamics studies of twist-stretch coupling under
the force shows that the intrinsic changes are similar to
$\mathcal{B-A}$ transition with much more smaller amplitude
\cite{Ref26}. Indeed, if unwouning of B-DNA tends to occur trough
A-form, decreasing of twist accompany with inclination and
decreasing of length. Also recent work \cite{RefBao17} suggests
that A- and B-forms helical structure of dsRNA and dsDNA are
responsible for difference in sign of twist-stretch coupling due
to additional inclination in A-form.

The switching mechanism between DNA conformations is sugar ring,
that can transform only though specific pathway. In the present
work the detailed study of coupling between elastic components and
specificity of coupling with internal component due to the
intrinsic flexibility of double helix (mainly associated with
sugar ring) has been performed. The approach
\cite{Ref27,Ref28,Ref29} is extended for tree coupled component
(two elastic and one intrinsic). We argue that observed changing
in the sings of coupling of elastic components under the force
could be interpreted due to coupling with intrinsic component as
rivalry to force influence. Similar behavior could be realized
between other couples of the deformation components. The obtained
quantitative results in framework of developed approach are in
agreement with the experiment data on single-molecule studies of
DNA mechanics and molecular modelling \cite{Ref26,Ref26a}.

\section{Model of DNA deformation.}
Let us consider the role of the coupling between deformation
components of macromolecules such as DNA. Deformation components
of elastic rod such as bend, twist and stretch are mostly
associated with local base pair step parameters as shift and/or
roll, twist and stretch respectively (fig.1). Deformation energy
of macromolecule in the model of an elastic rod is as follows:
\begin{equation}
\begin{aligned}
E_{WLC}=\frac{1}{2}\sum_{n=1}^{n=N}\{C_{R}(R_{n+1}-R_{n})^{2}+\\
C_{\varphi}(\varphi_{n+1}-\varphi_{n})^{2}+
C_{z}(Z_{n+1}-Z_{n})^{2}\},
\end{aligned}
\end{equation}
where $N$ is the number of monomer links of the molecular chain.
$R_{n}$ is dimensionless displacement of $n$- th link as whole
from the equilibrium position (the direction of displacement is
orthogonal to helix axis), $\varphi_{n}$ - turn of the $n-th$ link
as whole round the helix axis which is measured in radian, $Z_{n}$
- dimensionless displacement of $n$- th link as whole along helix
axis of macromolecular chain. $C_{R}, C_{\varphi}, C_{z}$ are
bending, twist and stretching rigidity of the macromolecule chain,
respectively.

The bending rigidity is related to the persistence length
$P={C_{R}l}/k_{B}T$, where $l=L/N$ - the average value of the link
length $l\sim 0,34 nm$, $k_{B}$ is the Boltzmann constant, $T$ is
temperature. According to various experiments bending rigidity is
$C_{R}=(85 \div 100) ~ kcal / mol$, that is typical for angle
about $ 6^{o}$\cite{Ref23} between normals to plan base pairs of
adjacent links and/or angel associated with shift. Torsion
rigidity constant is defined less precisely and strongly depends
on the sequence of base pairs, $C_{\varphi}\approx (64 \div 170) ~
kcal/mol$. Thermal deviation of torsion between adjacent  base
pairs of DNA is in the range of $(3.4^o \div 5.5^o)$ \cite{Ref20}.
Stretching rigidity, $C_{z}\sim S \cdot l=50~kcal/mol$ where $S
\sim 10^{3}~ pN$ is the stretch modulus of dsDNA or thermal
deviation $(0.02 \div 0.04) nm$ \cite{Ref24} .

\begin{figure}
\begin{center}
\includegraphics[height=39mm]{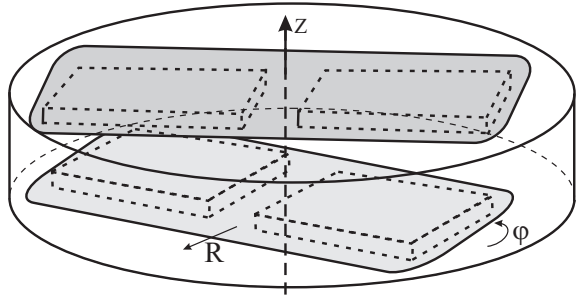}
\caption{Schematic image of the mobility base pairs as a whole
(link of molecular chain) in the elastic rod model.R - shelf of
coordinate relating with bend, $\varphi$ - twist coordinate,
$z$-stretch coordinate} \label{Compon}
\end{center}       
\end{figure}

In the model WLC  \cite{Ref1,Ref25} all the deformation components
vary independently. However, for the large amplitudes of
macromolecule deformation, coupling between components can provide
a significant contribution to the total energy of the system. Lets
examine the linear coupling of two of any elastic components of
the elastic rod ( $u, v =\{R, \varphi, Z\}$). In this case the
potential energy of the deformation can be written as:
\begin{equation}
\begin{aligned}
E=\frac{1}{2}\sum_{n=1}^{n=N}\{C_{u}(u_{n+1}-u_{n})^{2}+
C_{v}(v_{n+1}-v_{n})^{2}+\\
+2\gamma_{uv}(u_{n+1}-u_{n})(v_{n+1}-v_{n}),
\end{aligned}
\end{equation}
where $C_{u},C_{v}=\{ C_{R}, C_{\varphi}, C_{z}\}$, are the
rigidity constants of the macromolecule,
$\gamma_{uv}=\{\gamma_{R\varphi},\gamma_{Rz},\gamma_{z\varphi}\}$,
($\gamma_{uv}=\gamma_{vu}$) are the coupling parameter between $u$
and $v$ components.

Let us consider uniform deformation of the macromolecule chain.
Assume $u_{n+1}-u_{n}=u$, $v_{n+1}-v_{n}=v$. Introduce the
variables of the deformation of the chain is uniform, the energy
will be splitted into $N$ separated terms. Taking this in
consideration and introducing, the density of energy (2) in new
variables has the following form:
\begin{equation}
\varepsilon(u,v)=\frac{E}{N}=\frac{C_{u}u^{2}}{2}+\frac{C_{v}v^{2}}{2}+\gamma_{uv}uv.
\end{equation}
The energy (3) consist of two parts
$\varepsilon(u,v)=\varepsilon_{0}(u,v)+\varepsilon_{corr}(u,v)$,where
harmonic part
$\varepsilon_{0}(u,v)={C_{u}u^{2}}/{2}+{C_{v}v^{2}}/{2}$, and
correlated part $\varepsilon_{corr}(u,v)=\gamma_{uv}uv.$ We can
find minimum of the energy (3) as follows, in the other words it
is equations of stationary state:
\begin{equation}
\frac{\partial\varepsilon(u,v)}{\partial u}=C_{u}u+\gamma_{uv}v=0;
\end{equation}
\begin{equation}
\frac{\partial\varepsilon(u,v)}{\partial v}=C_{v}v+\gamma_{uv}u=0.
\end{equation}

The system has antiviral decision in the case of zero determinate
of coefficients. It gives following circumstance:
$C_{u}C_{v}-\gamma_{uv}^2=0$, then $v=v_a$ is any deformation, $u$
is determined  proportionally to $va$:
\begin{equation}
u=-\frac{\gamma_{uv}v_{a}}{C_{u}}
\end{equation}
However, in the case when $C_{u}C_{v}-\gamma_{uv}^2\neq 0$ and
$\gamma_{uv}=\gamma_{0}\sqrt{C_{u}C_{v}}$ and
$(-1\leq\gamma_{0}\leq1)$, stationary state is determined by zero
deviations from equilibrium state. Minimum of energy for
unstressed state of the system with coupling stays the same as for
system without coupling and realize for $u_{0}=0, v_{0}=0$,
$E_{0}=E(u_{0},v{0})=0$. In the same time minimum of energy will
be shifted under an external action. As a result uniformly
stressed state of chain with $v=v_{a}$ will have minimum of energy
in $u=u_{min}\neq 0$ due to coupling. For any deformations
according to the equations (4,5) minimum of the energy realized
when strain in both components are proportional to each other
$u(v)=-\gamma_{uv}v/C_{u}$, $v(u)= -\gamma_{uv}u/C_{v}$. Note,
that the positive sign in one of the components is accompanied by
a negative one in the other, in case $\gamma_{uv}>0$. There is the
same sign for both components in the case $\gamma_{uv}<0$.

Lets consider coupling in the next general view
$\gamma_{uv}=\gamma_{0}\sqrt{C_{u}C_{v}}$ and
$(-1\leq\gamma_{0}\leq1)$. Deviation from the equilibrium state
(or static strain) at one of the components, leads to a
proportional deformation in the other component, and the energy of
deformation (3) becomes:
\begin{equation}
\varepsilon(u,v_{a})=\frac{1}{2}C_{u}(u-u_{min})^2+\varepsilon_{a},
\end{equation}
where $u_{min}(v)=-\gamma_{0}\sqrt{C_{v}/C_{u}}v$,
$\varepsilon_{a}=\varepsilon(u_{min},v_{a})=1/2(1-\gamma_{0}^{2})C_{v}v_{a}^{2}$.
\begin{figure}
\begin{center}
\includegraphics[height=50mm]{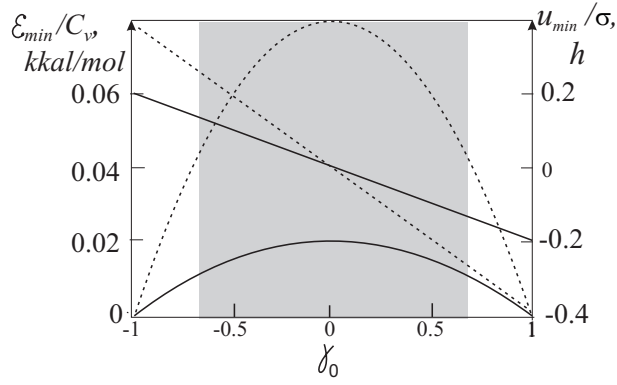}
\caption{Shift of deformation energy minimum in u-component in
response to deformation in v-component $v=v_{a}$. Small
interrelations $|\gamma_{0}|<\sqrt{1/2}$ correspond to grey area.
Large interrelations correspond to $\sqrt{1/2}<|\gamma_{0}|<1$
($v_a=0.2$ for solid lines, and $v_a=0.4$ for doted lines);
$\sigma=\sqrt{C_{u}/C_{v}}$; maximums of energy correspond to
$v_{a}^2/2$.} \label{profen}
\end{center}       
\end{figure}
The fig.2 shows the diagrams for two-components coupling in
depends on value of $\gamma_{0}$ in the range $[-1;1]$. The
diagrams show that modulus of coupling increasing leads to
increasing of u-component minimum shift value with decreasing of
the energy.

Since the DNA macromolecule is spiral and linear (generally
pre-curved), a sign for additional positive or negative strain in
u-component means the value of increase or decrease relatively
equilibrium value. Thus, when $\gamma_{0}>0$ a positive strain in
the v-component reduces the equilibrium deformation in the u
component through coupling. In other words, the positive
deformation in the u-component is complicated due to coupling with
a positive strained v-component. And in case of $\gamma_{0}<0$, a
positive strain in the v-component increases the equilibrium
deformation in the u component. Or a negative coupling constant
facilitates further positive straining the u-component.

Let us find the certain value $u=u_{c}$, for which deformation
energy u-component is equal to deformation energy on the same
value due to coupling with v-component. The value  $u_{c}$ is
found from the condition
$\varepsilon_{0}(u_{c},0)=\varepsilon(u_{c},v)~
(C_{u}u_{c}^2=C_{u}u_{c}^2+C_{v}v^2+2\gamma_{uv}v_{a}u_{c})$. Then
it has view $u_{c}=- {C_{v}v}/{\gamma_{uv}}$. There are 3 modes of
behavior depending on the relative position
 $u_{min}$ and  $u_{c}$. The first mode is
implemented when $|u_{min}|>|u_{c}|$, that corresponds to value of
coupling constant $|\gamma_{0}|>\sqrt{1/2}$. For the first mode
elastic one-component deformation energy on value $u_{min}$ is
higher then two-component deformation energy  on the same value
due to coupling with another deformed component on value $v_{a}$
and the inequality $\varepsilon_{0}(u_{min},0)>\varepsilon_{a}$ is
true. The second mode corresponds to the condition
$|u_{min}|=|u_{c}|$ or
$\varepsilon_{0}(u_{min},0)=\varepsilon_{a}$ and value
$|\gamma_{0}|=\sqrt{1/2}$. The third mode is in
$|u_{min}|<|u_{c}|$, and meets the condition
$|\gamma_{0}|<\sqrt{1/2}$. For the third mode the inequality
$\varepsilon_{0}(u_{min},0)<\varepsilon_{a}$ is true (Fig.3). In
this mode, two-component model with coupling probably realized
only for deformations which are larger then value $u_{c}$.

\begin{figure}
\begin{center}
\includegraphics[height=40mm]{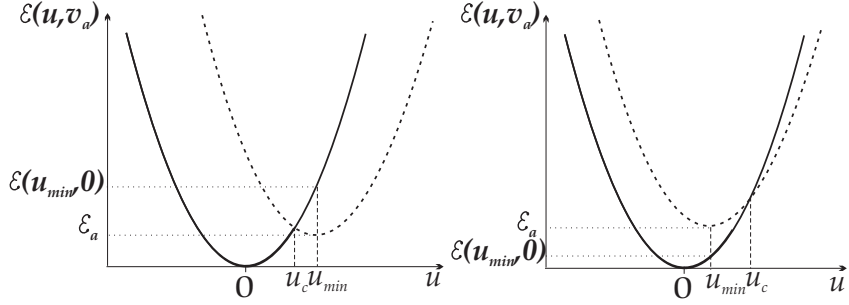}
\caption{Shift of deformation energy minimum in u-component in
response to deformation in v-component $v=v_{a}$. Large
interrelations correspond to $\sqrt{1/2}<|\gamma_{0}|<1$. Small
interrelations $|\gamma_{0}|<\sqrt{1/2}$ correspond to
$\varepsilon_{0}(u_{min},0)<\varepsilon_{a}$.
$\varepsilon_{0}(u_{min},0)>\varepsilon_{a}$} \label{profen}
\end{center}       
\end{figure}
The grey area shows the interval where coupling of component
correspond to the third mode. For this mode the interrelation
between components can lead to effective decrease one of
components rigidity. Replacing v-component of the expression
through u  in energy (3), the energy have the form:
\begin{equation}
\varepsilon(u)=\frac{1}{2}(C_{u}-\frac{\gamma_{uv}^2}{C_v})u^2
\end{equation}
Thus, the stiffness in one of the components can be effectively
reduced because the interaction between components.

The effective reducing is small to do noticed impact to character
of an elastic deformation. However, studying influence of
conformational rearrangement on bend of the chain we have
concluded that conformational transition can be advantageous in
deformed molecular chain of a DNA type in the case of the strained
fragment rigidity is lower then the average rigidity \cite{Ref28}.
So coupling between components can provide more probable
conformational transformations. The latter is probably mechanism
that couples elastic components.
\section{The deformation related with internal conformation change under the force}
Let us consider the deformations of the double helix, which are
accompanied by conformational changes in the structure of monomer
links. The potential energy of monomer links may be presented as a
non-linear function, $\Phi(r_n)$ of relative displacement of their
structural elements, $r_{n}$\cite{Ref27,Ref28,Ref29}. The variable
$r_{n}$ is different from elastic components, because the elastic
components describe relative motion of monomer links as whole
(external components). In the same time relative displacement of
structural elements inside a monomer link is described by internal
component $r_{n}$. The potential energy of the conformational
transformation according to trajectory determined by $\Phi(r_n)$
in the case when one external component is coupled with the
internal component may be presented as:
\begin{equation}
\begin{aligned}
E_{pot}=\frac{1}{2}\sum_{n=1}^{n=N}\{C_{v}(v_{n+1}-v_{n})^{2}+
C_{r}(r_{n+1}-r_{n})^{2}+\\
+\chi(v_{n+1}-v_{n-1})F(r_{n})+\Phi(r_{n})\},
\end{aligned}
\end{equation}
where  $~C_{v},~C_{r}$ are rigidity of external and internal
component along molecule correspondingly. $\chi F(r_{n})$
describes coordinated mobility of the external and internal
components. In the state of homogeneous conformation (all monomers
have equal $r_{n}=r$) density of potential energy can be written
like (4):
\begin{equation}
\varepsilon(v,r)=\frac{C_{v}v^2}{2}+ \chi vF(r)+\frac{\Phi(r)}{2},
\end{equation}
where $v_{n+1}-v_{n}=v$ is a change in one component of the
elastic rod,  $r_{n+1}-r_{n}=0$. In the minimum of conformational
energy $F(r)\sim-r$,$\Phi(r)=0$ or $const$, the energy (9) takes
the form similar to (3). However, in case of a local
conformational change (e.g. under the force action, in the
presence of proteins, intercalations with small molecules or site
specific conformational shift), the change of the elastic
component associated with conformational component is also
localized and proportionate to $F(r)$:
\begin{equation}
v=-\frac{\chi F(r)}{C_{v}}.
\end{equation}

To include in our consideration the mechanism allowing one of the
component changes,  applied  to one of  the components force is
considered. But it can't explain why sign of coupling parameter
changes under applied force about 35pN \cite{Ref26a,Ref26,Ref33}.
Based on our previous research, we can conclude that adding of
conformational component was missing link in those approaches.
Lets consider chain of two coupled components. One of them is
deformed by applied force and another coupled with generalized
conformational component. To isolate the role of component
coupling regards the case of small conformational changes $r
\rightarrow 0$. In this case,$\Phi(r) \rightarrow 0$, $\chi F(r)
\rightarrow -\chi_r$, the energy gets a view:
\begin{equation}
\varepsilon(u,v)=\frac{C_{u}u^{2}}{2}+\frac{C_{v}v^{2}}{2}+\gamma_{uv}uv-ufh-\chi_{r}v.
\end{equation}

To determine the ground state of the macromolecular chains, with
interrelation of components, we find the minimum of energy (4):
\begin{equation}
\frac{\partial\varepsilon(u,v)}{\partial
u}=C_{u}u+\gamma_{uv}v-fh=0;
\end{equation}
\begin{equation}
\frac{\partial\varepsilon(u,v)}{\partial
v}=C_{v}v+\gamma_{uv}u-\chi_{r}=0.
\end{equation}

Putting u from the first equation in to the second find
expressions for both components in the view:
\begin{equation}
v=\frac{\chi_{r}C_{u}-\gamma_{uv}fh}{C_{u}C_{v}-\gamma^{2}_{uv}};
\end{equation}
\begin{equation}
u=\frac{fh C_{v}-\gamma_{uv}\chi_{r}}{C_{u}C_{v}-\gamma^{2}_{uv}}.
\end{equation}
Both components have the same denominator, so the signs of the
components is determined by the numerators. The same sign of both
components, that corresponds effectively negative sign of coupling
component:
\begin{equation*}
 \begin{cases}
   \chi_{r}C_{u}-\gamma_{uv}fh>0,
   \\
  fh C_{v}-\gamma_{uv}\chi_{r}>0.
 \end{cases}
\end{equation*}
And different sing of components, that corresponds effectively
positive sign of coupling component:
\begin{equation*}
 \begin{cases}
   \chi_{r}C_{u}-\gamma_{uv}fh>0,
   \\
  fh C_{v}-\gamma_{uv}\chi_{r}<0.
 \end{cases}
\end{equation*}

Solving the systems of inequalities we find the circumstances,
which determine the signs of coupled components:
\begin{equation}
\frac{\gamma_{uv}}{C_{u}}<\frac{\chi_{r}}{fh}<\frac{C_{v}}{\gamma_{uv}},
\end{equation}
if components have the same sign.

\begin{equation}
\frac{\chi_{r}}{fh}<\frac{\gamma_{uv}}{C_{u}}<\frac{C_{v}}{\gamma_{uv}},
\end{equation}
if the components have different signs. So we have found the
circumstances between conformational parameter and force. The
circumstances determines signs of coupled components. In the first
case the system has solution the same sing that can be interpreted
as negative sign of coupling. In the second case, the solution
correspond to positive of coupling without conformational
component. This circumstance is in agreements with experiment
which demonstrate that under external force  the signs of twist
and stretching are the same, if the force value is lower then a
critical value. But for the force beyond the critical value, the
sings of twist and stretching are opposite.

\section{Discussion and Conclusions}
In the order to estimate the effect of conformational component on
twist-stretching coupling consider rigidity parameters in
accordance to experimental data \cite{Ref26a}. In the framework of
our model twist rigidity $C_{v}=197~kcal/mol$, stretching rigidity
$C_{u}=55~kcal/mol$, the twist-stretch coupling
$|\gamma_{uv}|=13.5~kcal/mol$, it correspond to the value of
$\gamma_{0}=0.13$. Component of stretching is extension deviation,
twist is rewriting as $\frac{\Delta Tw}{Tw_0}\approx\frac{v}{36^o
\pi/180^o}$. In the table we present values of deformations and
energy for different meanings of $\gamma_{0}$.
\begin{table}
\caption{Values of shift and energy minimum for stretching coupled
with twist}
\label{tab:0}       
\begin{tabular}{llll}
\hline\noalign{\smallskip} ~ &~~~ $\gamma_0$ & ~~~~~~$|u_{min}|$&~~~~~~$\varepsilon(u_{min},v_0)$ \\

\noalign{\smallskip}\hline\noalign{\smallskip}
              ~& ~~~ 0.13 & ~~~~~~0.003 & ~~~~~~0.016 \\
~$v_{0}=0.013$ &~~~ 0.3 & ~~~~~~0.007 & ~~~~~~0.015 \\
               &~~~ 0.7 & ~~~~~~0.017 & ~~~~~~0.008 \\
              ~& ~~~ 0.9 & ~~~~~~0.022 & ~~~~~~0.003 \\
\noalign{\smallskip}\hline
               ~ & ~~~ 0.13 &~~~~~~ 0.02&~~~~~~0.62 \\
 ~$v_{0}=0.08$& ~~~ 0.3 &~~~~~~ 0.045 & ~~~~~~0.574 \\
                & ~~~ 0.7 &~~~~~~ 0.106 & ~~~~~~0.322 \\
              ~& ~~~ 0.9 & ~~~~~~ 0.136 & ~~~~~~0.12 \\
\noalign{\smallskip}\hline
\end{tabular}
\end{table}
The values coincide with experimental data \cite{Ref26a} for
$\gamma_{0}=0.13$. Increasing of $\gamma_{0}$ leads to increasing
deformation \cite{Ref26a} and decreasing of energy.

Note the special case, when $\gamma_{uv}=\pm\sqrt{C_{u}C_{v}}$.
The density of energy may be written in the form:
\begin{equation}
\varepsilon(u,v_{0})=\frac{1}{2}C_{u}(u-u_{min})^2,
\end{equation}
where $u_{min}(v)=\mp\sqrt{C_{v}/C_{u}}v_0$. In this case
deformation in u-component induced by coupling with v-component
realized without additional energy.

In the case of force induced deformation with interrelation
between components of deformation and conformational changes we
estimate range of force where both coupled elastic components are
the same sign.
\begin{equation}
\frac{\chi_{r}\gamma_{uv}}{C_{v}}<fh<\frac{\chi_{r}C_{u}}{\gamma_{uv}};
\end{equation}
Substituting the values of the parameters in the relation (19),
the estimation for coupling function between conformational
component and twist is derived. The range of force of the same
sing of twist and stretch is:
\begin{equation}
0.07\chi_{r}<fh<4\chi_{r}.
\end{equation}
Since the maximum value of the force for such a regime is
$f_{c}=35 pN$ then last equation gives upper bound for
conformation - twist coupling ${\chi_{r}}/{h}< 8.8 pN$. As force
is more then $2pN$ rotation translates into twist changing
\cite{Ref26}, that determine low bound for conformation - twist
coupling ${\chi_{r}}/{h}>2.9 pN$. Hence while the double helix
conformation resists to force influence, the deformation occurs in
the specific way. As ever double helix undergoes conformational
changes  under the force the deformation goes in the other way.
Parameter $\chi_{r}$ is a barrier which is formed by
conformational stability of $v-$component in the range of minima
of pseudorotation angle.

Thus, the paper presents a model of DNA deformation that takes
into account the internal conformational component to describe
coupled elastic deformations. Due to the coupling of the elastic
component with the conformational component, additional term
arises in the coupled elastic components solutions.  This term
competes with the contribution associated with the directed action
of the applied force and provides possibility of change of
character of deformation. Thus, the model not only allows one to
describe the unusual relationship between twisting and stretching
of a macromolecule under the action of an applied force, but also
to clarify its nature contained in the features of the internal
structure of the double helix.

\end{document}